\documentclass[superscriptaddress,twocolumn,showpacs,preprintnumbers,amsmath,amssymb]{revtex4}

\usepackage{graphicx}
\usepackage{dcolumn}
\usepackage{bm}

\begin{document}

\preprint{}

\title{Specific Heat of the $S=1/2$ Two-Dimensional Shastry-Sutherland
Antiferromagnet \\ SrCu$_2$(BO$_3$)$_2$ in High Magnetic Fields}

\author{H. Tsujii}
\affiliation{Department of Physics, University of Florida, P. O.
Box 118440, Gainesville, Florida 32611-8440}
\author{R. C. Rotundu}
\affiliation{Department of Physics, University of Florida, P. O.
Box 118440, Gainesville, Florida 32611-8440}
\author{B. Andraka}
\affiliation{Department of Physics, University of Florida, P. O.
Box 118440, Gainesville, Florida 32611-8440}
\author{Y. Takano}
\affiliation{Department of Physics, University of Florida, P. O.
Box 118440, Gainesville, Florida 32611-8440}
\author{H. Kageyama}
\affiliation{Institute for Solid State Physics, University of
Tokyo, Kashiwa, Chiba 277-8581, Japan}
\author{Y. Ueda}
\affiliation{Institute for Solid State Physics, University of
Tokyo, Kashiwa, Chiba 277-8581, Japan}

\date{\today}

\begin{abstract}
We characterize the field-induced magnetic phases of
SrCu$_2$(BO$_3$)$_2$, a frustrated spin-1/2 Heisenberg
antiferromagnet in the two-dimensional Shastry-Sutherland lattice,
using specific heat in magnetic fields up to 33 T.  We find that
the spin gap persists above the expected critical field ${H_c =
\Delta/{\rm g}\mu_{\rm B}}$ of 21 T despite the appearance of
magnetic moment in the ground state.  At the magnetization plateau
at 1/8 of the saturation, the ${S_z = +1}$ triplets that carry the
magnetization of the ground state are observed to form a
two-dimensional spin gas of massive bosons.  A spin gas consisting
of the same number of massive particles continues to completely
dominate the specific heat in the field region above the plateau,
although the magnetization increases with increasing field.
Ordering is observed at a temperature immediately below the
spin-gas regime.
\end{abstract}

\maketitle

Geometrical frustration is one of the important subjects in
strongly correlated systems \cite{Ram}. In particular, the
fascinating interplay of geometrical frustration and quantum
fluctuation in two dimensions provides a fertile ground for new
physics, as has been theoretically demonstrated for Heisenberg
antiferromagnets in the triangular lattice and the kagom\'{e}
lattice.  For the $S = 1/2$ triangular antiferromagnet, there now
appears to be a consensus that N\'{e}el order must exist despite
the strong frustration \cite{Trig}.  The $S = 1/2$ kagom\'{e}
antiferromagnet has been predicted to possess a gap separating the
ground state from upper triplet levels and a band of low-lying
singlet excitations within the triplet gap, although the exact
nature of the ground state is still under debate \cite{Kag}. These
predictions are yet to be borne out by experiment, because no
isotropic $S = 1/2$ model system has been identified in the
laboratory for these lattices.

In the last few years, there has been a rapid progress in the
understanding of the $S = 1/2$ Heisenberg antiferromagnet in the
Shastry-Sutherland lattice \cite{SS}, another two-dimensional
frustrated geometry, for which the exact ground state is known at
zero and low magnetic fields despite geometrical frustration.  The
progress owes largely to the discovery of the spin gap and other
novel magnetic properties in SrCu$_2$(BO$_3$)$_2$ \cite{1}, the
only known laboratory model for the geometrically frustrated $S =
1/2$ two-dimensional Heisenberg antiferromagnet.

The low-field ground state of the Shastry-Sutherland
antiferromagnet is simply a product of spin-dimer singlet wave
functions, and the dimerization by the diagonal bond $J$ causes a
gap to form \cite{SS}.  The gapped triplet excitations, which are
also spin dimers at least to a good approximation, have a large
mass as a result of the unique lattice topology involving
orthogonal arrangement of dimers \cite{Miya, Zheng2, MH, Miya2}.
In SrCu$_2$(BO$_3$)$_2$, the coupling strengths extracted from
experiments are $J=85{\rm ~K}$ for the intra-dimer exchange and
$J'=54{\rm ~K}$ for the inter-dimer exchange  \cite{Miya3} or
$J=70{\rm ~K}$ and $J'=42{\rm ~K}$ \cite{Knet}. The large $J'$
makes the energy gap $\Delta$ substantially smaller than $J$
\cite{1, C}.  In fact, the ratio $J'/J$ is very close to the
critical value 0.70 \cite{Miya, MH, Zheng} for the quantum phase
transition to a square-lattice N\'{e}el state.  In other words,
$J$ and $J'$ are extremely frustrated.

The striking feature of SrCu$_2$(BO$_3$)$_2$ is the existence of
magnetization plateaus at 1/8, 1/4, and 1/3 of the saturation
magnetization \cite{1,2}.  Although there is a substantial
spin-phonon coupling in this compound \cite{Phon}, the plateaus
are an intrinsic property of Heisenberg spins in the
Shastry-Sutherland lattice and due to geometrical frustration
\cite{Miya, Mis}.  The spin state at any magnetization plateau is
obviously gapped \cite{Oshi}, and a possible connection between
magnetization plateaus and quantized Hall-conductance plateaus has
been pointed out \cite{Mis}.  In contrast, the states between
$H_c$ and the first plateau, between two adjacent plateaus, and
between the highest plateau and the saturation field have been
believed to be all gapless by a prevalent view on magnetization
plateaus.

    In this paper, we examine three field-induced magnetic
phases of SrCu$_2$(BO$_3$)$_2$: the one between $H_c$ and the
first plateau with 1/8 of the saturation magnetization, the
plateau phase, and much of the phase between the 1/8 and 1/4
plateaus.  The results refute the conventional view on the nature
of the first phase and uncover the presence of a two-dimensional
spin gas in the other phases, as well as ordering in the third
phase.

    Prior to our work, two-dimensional Bose gas has been realized
in spin-polarized atomic hydrogen adsorbed on the surface of
superfluid $^4$He \cite{Mosk}.  However, the spin gas in
SrCu$_2$(BO$_3$)$_2$ is the first example that is amenable to
thermodynamic studies.

The calorimeter for this investigation employed the relaxation
method \cite{Rel}.  In order to optimize the relaxation time at
different temperature and magnetic-field regions, three samples of
different sizes were used, all taken from a high-purity single
crystal whose growth procedure has been described in Ref.
\cite{X}.  The sample sizes were roughly
3~$\times$~3~$\times$~0.07~mm$^3$,
4~$\times$~3~$\times$~0.3~mm$^3$, and
3~$\times$~3~$\times$~1.3~mm$^3$, and the magnetic field was
applied parallel to the crystalline {\em c} axis, perpendicular to
the two-dimensional layers of the $S = 1/2$ Cu$^{2+}$ spins.  In
this field orientation, the spin gap for the $S_z = +1$ triplets
is split into two branches that are 4.1 K apart \cite{ESR, ESR2}
by the Dzyaloshinskii-Moriya interaction \cite{DM, Neutr}.
According to the {\sl g} factor measured by ESR \cite{ESR}, the
lower gap is expected to close at ${H_c = \Delta/{\rm g}\mu_{\rm
B}}$ of 21 T.

The measurements were done primarily in magnetic fields between 22
T and 33 T at temperatures between 0.53 K and 15 K.  Additional
data were taken at zero field and 14 T for comparison with earlier
specific-heat results at magnetic fields up to 12 T \cite{C}.  At
zero field, there was excellent agreement between the present data
and the earlier result, indicating a high sample quality and
reproducibility.  The 14 T data precisely follow the field
dependence observed in the earlier experiment, with the spin gap
decreasing linearly with magnetic field and the position of the
broad maximum remaining at 8 K.

Figure 1 shows the specific heat at 22 T and 24~T, where the spin
gap is expected to be absent and the ground state is magnetic
\cite{1,2}, together with the 14~T result for comparison.  The
$T^3$ phonon contribution to the specific heat has been subtracted
from all the data presented in this paper, as determined by
Kageyama {\em et al}. \cite{C}  Instead of a broad maximum at 8~K,
a new hump appears at 22 T and 24 T at temperatures below 1.4~K.
The position of this low-temperature maximum decreases with
increasing magnetic field.  At 22 T, the low-temperature tail of
the hump clearly shows a gapped behavior, whereas the 24 T data do
not extend to a sufficiently low temperature.

\begin{figure}[btp]
\begin{center}\leavevmode
\includegraphics[width=0.8\linewidth]{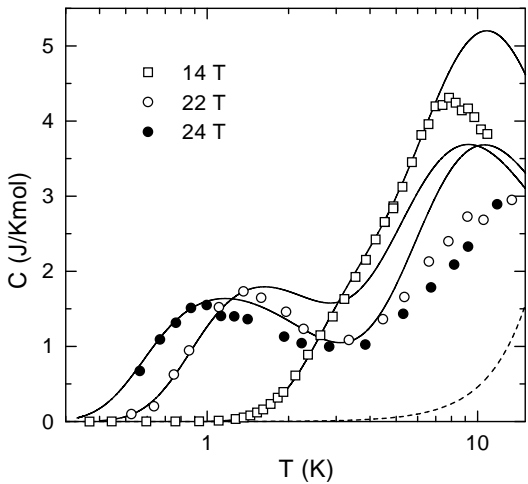}
\caption{ Specific heat of SrCu$_2$(BO$_3$)$_2$ at 14 T, which is
below the expected critical field $H_c$ of 21 T, and at 22 T and
24 T in the region between $H_c$ and the first magnetization
plateau at 1/8 of the saturation.  The solid lines are the fits
discussed in the text.  The broken line is the $T^3$ phonon
contribution that has been subtracted from the
data.}\label{fig1}\end{center}\end{figure}

The overall behavior of the specific heat can be explained in
terms of the dimer-quadrumer model of Hofmann {\em et al}.
\cite{Hof}, who incorporated the repulsion between triplet spin
dimers and bound pairs of triplet dimers as a mean field. The line
for 14 T in the figure is a calculation using exactly the same
parameters chosen by these authors to fit the specific-heat data
at fields up to 12 T.  Similar to their fits, the 14 T calculation
reproduces the experimental data very well at temperatures below
the broad peak.

In order to fit the data at 22 T and 24 T, we have found that it
is necessary to allow the gap energy $\Delta_+$ of the $S_z = +1$
triplets and the mean-field strength $v_0$ of the repulsion term
to be fitting parameters \cite{V}.  The best fits shown in the
figure are obtained by choosing $\Delta_+ = 4.6{\rm~K}$ and 3.2~K
for 22 T and 24 T, respectively, and $v_0 = 10{\rm~K}$ for both
fields.  The fit at 22 T is good primarily at temperatures below
the broad low-temperature peak, whereas the fit at 24 T agrees
well with the data up to 4~K.  The good overall agreement leads us
to conclude that the ground state is gapped at both fields.

The remarkable persistence of the gap in the field-induced
magnetic phase contradicts the widely held view that the ground
state above $H_c$ evolves with increasing magnetic field by simply
increasing the number of $S_z = +1$ triplets within itself. We
propose in contrast that mixing of at least one of the $S_z = +1$
triplet levels with the singlet ground state is responsible for
the evolution of the magnetization below the 1/8 plateau and for
the anticrossing that keeps the gap open even above $H_c$. The
mixing is probably caused by the Dzyaloshinskii-Moriya
interaction.  The gap keeps the ground state in a spin-liquid
phase with a non-zero magnetization. This explains why the
translational symmetry is maintained in this field region even at
a temperature as low as 35 mK, as has been found by NMR \cite{N}.
We conclude that the region from zero field up to the lower edge
of the 1/8 magnetization plateau is a contiguous phase, with no
ordering even at fields above $H_c$.

The situation resembles the behavior of the Haldane-gap
antiferromagnet NENP, where the staggered {\sl g} tensor of two
non-equivalent Ni sites causes the triplet excitation to anticross
the ground state at $H_c$ \cite{Palme, End} and magnetization
develops with no long-range order due to the inter-chain coupling
\cite{Katsu}.  It is interesting to note that isolated triplet
spin dimers in the Shastry-Sutherland lattice have been predicted
to be unstable below the 1/8 plateau and they cannot be
constituents of the ground state in this field region \cite{F}.

The present result is consistent with the ESR experiments
\cite{ESR, ESR2}, which have found no excitation level that
reaches the ground state at $H_c = 21$ T.  In fact no such level
has yet been discovered up to 40 T.  Instead, the behavior of the
lower $S_z = +1$ branch, and possibly also the upper branch,
suggests anticrossing around $H_c$ \cite{ESR2}.

The fitting parameters are also in good agreement with the ESR
data \cite{ESR2}, which give $\Delta_+ = 4.8$ K and 6.6 K for the
two branches of excitations at 22 T.  The ESR values at 24 T are
5.2 K and 5.5 K.  Our value for the repulsion parameter $v_0 = 10$
K at 22 T and 24 T is considerably smaller than 17 K for fields up
to 12 T \cite{Hof} and at 14 T.  This is expected, since the
mixing of an $S_z = +1$ excited state with the ground state at
fields above $H_c$ turns it into a more singlet-like object, which
experiences less repulsion due to the inter-dimer exchange $J'$.

Figure 2 shows the specific heat for fields of 27.5~T and higher.
27.5~T corresponds to the midpoint of the 1/8 magnetization
plateau, where Cu NMR has found evidence for a rhomboid ordered
structure involving $S_z=+1$ triplets \cite{N}.  Within our
experimental accuracy, the specific heat at this field is 1/8 of
the gas constant $R$ over a wide temperature range except for a
gradual rise at temperatures above 5 K and a small decrease at
temperatures below 1 K.  This is conclusive evidence that a
two-dimensional spin gas is formed by the $S_z = +1$ triplets
whose number equals 1/8 of the total spin dimers.  This conclusion
is consistent with the $^{11}$B NMR spectra \cite{N}, according to
which the spin superstructure disappears somewhere between 35 mK
and 1.5 K.  Evidently, the transition from a rhomboid spin solid
to the spin gas occurs at a temperature below 0.63 K, our lowest
temperature for the plateau region.

\begin{figure}[btp]
\begin{center}\leavevmode
\includegraphics[width=0.8\linewidth]{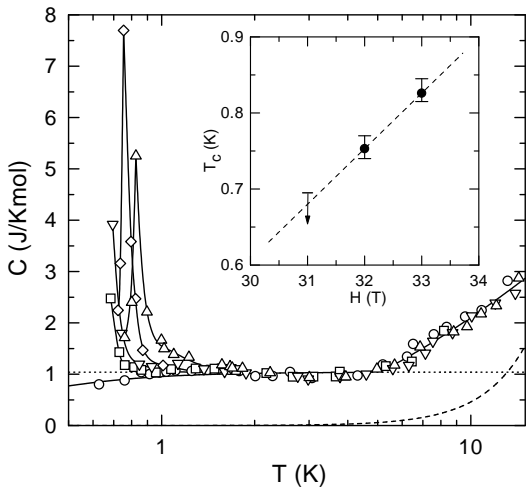}
\caption{ Specific heat of SrCu$_2$(BO$_3$)$_2$ at 27.5 T
($\circ$), which is the midpoint of the 1/8 magnetization plateau,
and at higher fields in the region between the 1/8 and 1/4
plateaus: 29.6~T ($\square$), 31~T ($\triangledown$), 32~T
($\diamond$), and 33 T ($\vartriangle$).  The dotted line
indicates $C=R/8$. The solid line through the 27.5 T data points
below 5 K is a single-parameter fit for a non-interacting
two-dimensional Bose gas.  Other solid lines are guides to the
eye.  The $T^3$ phonon contribution shown by the broken line has
been subtracted from the data.  The inset shows the ordering
temperature as a function of magnetic field.
}\label{fig2}\end{center}\end{figure}

The solid line drawn in Fig.~2 through the 27.5 T data points
below 5 K is the heat capacity of 1/8 mole of ideal
two-dimensional Bose gas \cite{BE}, with the particle mass being
the only fitting parameter. The fit yields a bosonic mass of
$(8.7\pm1.0)\times10^{-28}$ kg, which is 960 times the electronic
mass.  It is 62\% of the mass of the lowest triplet excitation,
$(1.4\pm0.1)\times10^{-27}$ kg, obtained from a sinusoidal fit of
the dispersion measured by inelastic neutron scattering at zero
field \cite{NeutrK}. The similarity of the two masses suggests
that the triplet excitations at lower fields are somewhat extended
objects \cite{N} similar to the ground-state triplets at the
magnetization plateau.  It is important to note that the thermal
de Broglie wavelength of the spin gas at 0.63 K is as large as 1.7
times the average interparticle distance. Hence, the freezing of
the spin gas into a rhomboid structure takes place in a highly
degenerate regime.

Surprisingly, the specific heat between 29.6~T and 33~T, in the
field region above the magnetization plateau, is completely
different from that in the region below the plateau and is
identical with the 27.5 T data except for the low-temperature
anomaly that signals phase transition.  In particular, it remains
at $R/8$ in the mid-temperature range within our experimental
accuracy, demonstrating that the same number of particles form a
two-dimensional spin gas at these fields as at the plateau. It is
natural to conclude that these particles at the two field regions
are identical.

On the other hand, the magnetization above the plateau increases
continuously with increasing field and reaches a 52\% higher value
at 33 T \cite{1, 2}. It is very unlikely that the extra
magnetization is carried by the particles of the spin gas, since
$\langle S_z \rangle$ will then have to be larger than unity and
this would require mixing of an $S_z = +2$ excited state.  A
presence of such a state should result in a low-temperature
specific-heat hump whose position changes with magnetic field,
very much like the behavior shown in Fig.~1 for the field region
below the magnetization plateau.  We propose that the extra
magnetization is carried by additional $S_z = +1$ particles that
are thermodynamically inert at least up to the highest temperature
of this experiment.

Phenomenologically, the particle-number conservation that
underlies the two-dimensional gas behavior can be explained by
large energy required for excitations that do not conserve the
particle number.  We believe, however, that this remarkable
phenomenon deserves a fundamental explanation, which is presently
lacking.

Momoi and Totsuka \cite{MT} have predicted that two species of
$S_z = +1$ bosons compose the state above the 1/3 magnetization
plateau, beyond the field region explored by experiments to date.
They are those same particles that form the plateau phase and
additional particles that condense into a magnetic superfluid,
whose density increases with increasing magnetic field.  A similar
two-species state in the field region above the 1/8 magnetization
plateau can explain why the number of particles that make up the
spin gas remains the same as in the plateau phase.  However, the
robustness of the particle number against thermal excitations
still remains a mystery.

The sharp transition marked by the low-temperature peak in the
specific heat is probably due to freezing of the spin gas
\cite{Nuc}.  The strong field dependence of the transition
temperature indicates that either the particle mass or the
strength of the repulsive interaction increases with the density
of the thermodynamically invisible particles.  Since freezing in
two dimensions is generally a Kosterlitz-Thouless transition
\cite{KT}, which has no signature in specific heat, the sharp peak
suggests an importance of spin-phonon coupling or inter-layer spin
coupling.

In summary, we have uncovered a distinct spin behavior in each of
three high magnetic-field regions of SrCu$_2$(BO$_3$)$_2$.  In the
region below that of the 1/8 magnetization plateau, where
translational symmetry exists, the spin gap remains open,
indicating hybridization of at least one of the two $S_z = +1$
triplet states with the singlet ground state, which now acquires a
magnetic moment.  We conclude that there is no phase transition
between this region and the low-field, non-magnetic region.  The
regions at and above the magnetization plateau constitute a second
contiguous phase, where those triplets whose number equals 1/8 of
the spin dimers form a two-dimensional gas of massive bosons above
the field-dependent freezing temperature.  In addition, the region
above the plateau contains thermodynamically inert particles that
carry the magnetization in excess of 1/8 of the saturation value.
Further studies are called for to identify these carriers and to
elucidate their role in the ordering.

We thank N.~Aso, T.~C.~Kobayashi, G.~Misguich, S.~Miyahara,
Y.~Narumi, M.~Nohara, H.~Nojiri, M.~Takigawa, N.~Tateiwa, and
G.~S.~Uhrig for useful communications and/or for providing data,
some prior to publication. We have benefited from stimulating
discussions with E.~Dagotto, S.~Haas, J.~S.~Kim, P.~Kumar,
D.~L.~Maslov, S.~P.~Obukhov, and M.~Takigawa.  The experiment
above 22 T was performed at the National High Magnetic Field
Laboratory, which is supported by NSF Cooperative Agreement
No.~DMR-0084173 and by the State of Florida.  We thank
B.~L.~Brandt, S.~T.~Hannahs, T.~P.~Murphy, and E.~C.~Palm for
their help and hospitality. This work was funded by the NSF
through DMR-9802050 and DMR-0104240 and by the MECSST (Japan)
through Grant-in-Aid for Scientific Research No.~40302640.


\begin{thebibliography}{99}
\bibitem{V}   The repulsive interaction between neighboring triplets is  $v =
v_0(n_1 + 3n_2/2)$, where $n_1$ and $n_2$ are the densities of
triplets and pairs of triplets, respectively.  See Ref. 25.
\bibitem{BE}  As is well known, the specific heat of the non-relativistic
ideal gas in two dimensions does not depend on whether the
particles are bosons or fermions.  Hence, no assumption needs to
be made on the statistics of the particles in comparing the data
with an ideal-gas behavior and extracting the particle mass.
\bibitem{Nuc} These anomalies are not spurious effects associated with the
nuclear specific heat, which comes mostly from $^{11}$B,
$^{63}$Cu, and $^{65}$Cu and is at most 0.03 J/Kmol at all fields
even at our lowest temperature.
\end{thebibliography}
\end{document}